\documentclass{article}
\usepackage[utf8]{inputenc}
\usepackage[margin=1in]{geometry}
\usepackage{times}
\usepackage{helvet}
\usepackage{courier}
\usepackage{color}
\usepackage{tikz}
\usepackage{hyperref}
\usepackage{graphicx}
\usepackage{tabularx}
\usepackage{float}
\usepackage{natbib}
\usepackage{amsmath}
\usepackage[utf8]{inputenc}
\usepackage{booktabs}
\usepackage{setspace}
\usepackage[utf8]{inputenc}
\usepackage{amsmath}
\usepackage{appendix}
\usepackage{booktabs}
\usepackage{graphicx}
\usepackage{caption}
\usepackage{url}
\usepackage{subcaption}
\usepackage{lineno}
\usepackage[export]{adjustbox}

\begin{document}
\title{Insidious Nonetheless: How Small Effects and Hierarchical Norms Create and Maintain Gender Disparities in Organizations\footnote{Our title comes from a quote by computer scientist Lenore Blum, who resigned from her university position as a consequence of cumulative bias: “Subtle biases and micro-aggressions pile up, few of which on their own rise to the level of ‘let’s take action,’ but are insidious nonetheless.” \citep{certoLenoreBlumShocked2018}}}

\author{Yuhao Du\textsuperscript{a}, Jessica Nordell\textsuperscript{b}, Kenneth Joseph\textsuperscript{a,1} \\ 
{\small \textsuperscript{a} Department of Computer Science and Engineering, University at Buffalo, Buffalo, NY, 14260} \\ 
{\small \textsuperscript{b} Author, \emph{The End of Bias: A Beginning}} \\
{\small \textsuperscript{1} Corresponding Author: kjoseph@buffalo.edu}}


\maketitle

\begin{abstract}
The term \emph{glass ceiling} is applied to the well-established phenomenon in which women and people of color are consistently blocked from reaching the upper-most levels of the corporate hierarchy. Focusing on gender, we present an agent-based model that explores how empirically established mechanisms of interpersonal discrimination coevolve with social norms at both the organizational (meso) and societal (macro) levels to produce this glass ceiling effect for women. Our model extends our understanding of how the glass ceiling arises, and why it can be resistant to change. We do so by synthesizing existing psychological and structural theories of discrimination into a mathematical model that quantifies explicitly how complex organizational systems can produce and maintain inequality. We discuss implications of our findings for both intervention and future empirical analyses, and provide open-source code for those wishing to adapt or extend our work.
\end{abstract}

Men are overrepresented at higher levels of the corporate hierarchy. The New York Times reports, for instance, that in 2018 there were fewer female chief executives at Fortune 500 companies than male chief executives with the name James, despite the fact that only 3.3\% of the U.S. population is named James, while women make up 50.3\% of the U.S. population  \citep{millerTopJobsWhere2018}. 

Scholars have long studied potential reasons for this \textit{glass ceiling effect} - the name given for the general phenomenon in which invisible barriers block women and people of color from reaching high levels of management \citep{cotterGlassCeilingEffect2001,bertrandCoaseLectureGlass2018}. Here, we focus on literature specifically surrounding gender. Some popular explanations of the glass ceiling revolve around innate or learned differences between men and women, such as psychological differences in risk-taking or taste for competition/negotiation \citep{schubertGenderSpecificAttitudes2000a,reubenTasteCompetitionGender2015,babcockWomenDonAsk2004}, or differences on personality traits \citep{filerSexualDifferencesEarnings1983,semykinaGenderDifferencesPersonality2007,collischonPersonalityTraitsPartial2021}. Others have focused beyond the individual, to the places where gender norms---roughly, culturally-prescribed guidelines for behavior, based on one's own perceived gender and the perceived gender of those around us--- and stereotypes---generalized and often unfounded assumptions about someone based on their (perceived) gender---are learned and enforced. To this end, scholars have found that policy, including family leave and flexible scheduling \citep{williamsMaternalWallRelief2003,pettitStructureWomenEmployment2005,goldinMostEgalitarianProfession2016,bearForgetMommyTrack2021}, and interpersonal factors such as harassment \citep{stockdaleSexualHarassmentGlass2009,berdahlWorkplaceHarassmentDouble2006} and gender-biased evaluations \citep{moss2012science,heilmanNoCreditWhere2005}, both play significant roles in creating or limiting the upward mobility of women in the workplace.

Due to the limits of what can be operationalized in a single study, efforts to empirically identify causes of the glass ceiling rarely consider more than a few competing ideas, and often do so at a single (or a few) moments in time. This can be problematic, because corporations are examples of \emph{complex social systems} \citep{harrison2007simulation,martellBiasExclusionMultilevel2012}, where social norms and stereotypes diffuse over time through individuals and groups within the organization, and back and forth between the organization and society. The interaction of these multiple and hierarchical social structures create feedback processes that can confound simple explanations of empirical findings. For example, a complex systems view of gender shows that empirical observations of individual-level differences based on gender can be explained by overarching cultural norms, without need to rely on tropes of biological or personality-based gender or sex differences \citep{ridgewayFramedGenderHow2011,markWhyNominalCharacteristics2009}.  Empirical work therefore cannot cleanly address the fact that in the real world, women experience gender discrimination in many ways, over long periods of time. 

Acknowledging the limitations of empirical work, scholars have turned to \emph{simulation}, and in particular, \emph{agent-based modeling}, to study gender disparities in organizations \citep{martellMalefemaleDifferencesComputer1996a,robison-coxSimulatingGenderStratification2007,bullinariaAgentBasedModelsGender2018,momennejadComputationalJusticeSimulating2019}. In an agent-based model, a computational, simplified representation of an individual (an “agent”) interacts with other agents using a predefined set of rules. These rules shape macro-level statistics, which can then “feed back” to reshape the parameters of the established rules \citep{gilbertAgentBasedModels2007}. Agent-based models have long been used in the social sciences to study phenomena within complex systems, because one can rapidly consider experiments that are too large for empirical study and can also easily examine counterfactual arguments within evolving systems  \citep{carleyTheoryGroupStability1991}. 

The present work proposes a new agent-based model of how the glass ceiling emerges within the complex social system of a hypothetical corporation. We outline how glass ceilings within organizations can emerge through a coupling of 1) stable, hierarchical gendered norms about who belongs where and has what value, and 2) small, discrete, empirically-validated instances in which these norms are enacted at the interpersonal level. We use this new model of the glass ceiling effect to study how assumptions about interpersonal discrimination, and its interactions with these hierarchically structured social norms, impact the success or failure of a quota-based intervention.

Our work extends knowledge of how the glass ceiling arises in two ways. First, prior work has focused largely on gender bias as a broadly defined phenomenon that occurs in organizations \citep{martellBiasExclusionMultilevel2012}, or as the product of a more general theory of social behavior, e.g. status characteristics theory \citep{ridgewayFramedGenderHow2011}. Here, we instead introduce a specific set of common and empirically demonstrated mechanisms in which gender discrimination occurs at the interpersonal level, and connect these to the dynamics of the organization. Doing so provides a critical link between empirical measures of interpersonal gender discrimination and the complex system in which they are embedded. Second, our model emphasizes a need to incorporate how both organizational and societal-level social norms are linked to gender discrimination within organizations. Taking cues from literature on racial inequality \citep{rayTheoryRacializedOrganizations2019}, we label organizational-level norms as \emph{meso-level norms}, and societal-level norms as \emph{macro-level norms}.  We describe how social norms at the meso- and macro-levels interact with discrimination at the interpersonal level in a feedback loop that reproduces gender disparities in a diverse array of organizations and that can restrict the impact of typical interventions in the long term.

\section{Model Overview}

\begin{table*}[t]
\small
    \centering
    \begin{tabular}{| p{.25\textwidth}|p{.7\textwidth}|}
    \toprule
{\bf Mechanism} & {\bf Model Implementation}  \\ \hline \toprule
{\bf Reward Individual Success} & Women receive a smaller increase in their perceived promotability when a project succeeds (\ref{sec:reward_indiv}) \\ \hline
{\bf Penalty Individual Failure} & Women receive a greater decrease to their perceived promotability when a project fails (\ref{sec:penalty_indiv}) \\ \hline
{\bf Reward Mixed Group Success} & Women receive a smaller increase in promotability when a  mixed-gender project succeeds (\ref{sec:reward_group}) \\ \hline
{\bf Penalty Mixed Group Failure} & Women receive a larger decrease in promotability when a  mixed-gender project fails (\ref{sec:penalty_group}) \\ \hline
{\bf Penalties for Non-Altruism} & Women will occasionally complain about unfairness when they experience bias. Doing so leads to a decrease in their perceived promotability (\ref{sec:altruistic}). \\ \hline
{\bf Penalty Stretch Project} &  Stretch projects are differentially assigned to men over women (\ref{sec:growth}) \\ \hline
    \end{tabular}
    \caption{The six empirically-validated interpersonal gender discrimination mechanisms that we add to the unbiased model. We provide a brief description of how each is implemented, and a link to the section of the text with full details on how the mechanism is implemented in our model. }
    \label{tab:bias_list}
\end{table*}


Our work is based on the seminal simulation model of \cite{martellMalefemaleDifferencesComputer1996a}, who show how gender disparities in the corporate hierarchy can arise solely from small gender biases during performance evaluation.   As in their work, agents in our model represent employees of a hypothetical, eight-level corporate organization, with a pre-specified number of agents at each level.  
The primary difference between our model and  \cite{martellMalefemaleDifferencesComputer1996a} is that we link gender disparity not to a generalized notion of "bias" in performance evaluations, but to specific empirically identified mechanisms through which this bias manifests. To do so, our model simulates two common process in organizations: employees engage in projects, and employees are promoted through the ranks of the company. Projects may succeed or fail, and promotions are based on the agents’ \emph{perceived promotability}. 

At the start of the simulation, agents are randomly initialized with a perceived binary gender (male or female) and a perceived promotability that we assign randomly. The simulation then iterates over a series of \emph{turns}. On each turn, agents receive either an individual or group \emph{project}. At fixed intervals, we also introduce stretch projects that provide outsized boosts in perceived promotability. This project then (randomly) succeeds or fails with equal probability, irrespective of the agent’s perceived promotability or gender. When an agent’s project succeeds, the agent receives some \emph{credit} that increases their perceived promotability. When a project fails, the agent’s perceived promotability drops via some amount of credit.

After some number of simulation turns, there is a \emph{promotion cycle}. During a promotion cycle, the employees with the highest perceived promotability move up from their current level of the corporate hierarchy to the next. To make room for them, a random 15\% of the individuals at each level of the hierarchy leave the organization. These spots are then recursively filled until the bottom of the hierarchy is reached. At this point, new agents are then created and “hired” into the entry level of the company. These new agents are equally likely to be men or women.

In this \emph{unbiased model}, there are no differences between men and women: they are equally likely to begin with a given level of promotability, to succeed and fail at projects, to receive stretch projects, and to leave the company. We introduce our model of how the glass ceiling arises through two experiments that extend this unbiased model. First, we introduce six specific, empirically-observed ways in which gender discrimination at the interpersonal level manifests in the workplace. Second, we propose a mechanism through which interpersonal gender discrimination is tied to gendered social norms at the macro and meso levels. 



\section{Results}

\subsection*{Incorporating Small Interpersonal Acts of Discrimination into an Unbiased Model}

\begin{figure}[t]
     \centering
     \begin{subfigure}[b]{\textwidth}
         \centering
         \includegraphics[width=\textwidth]{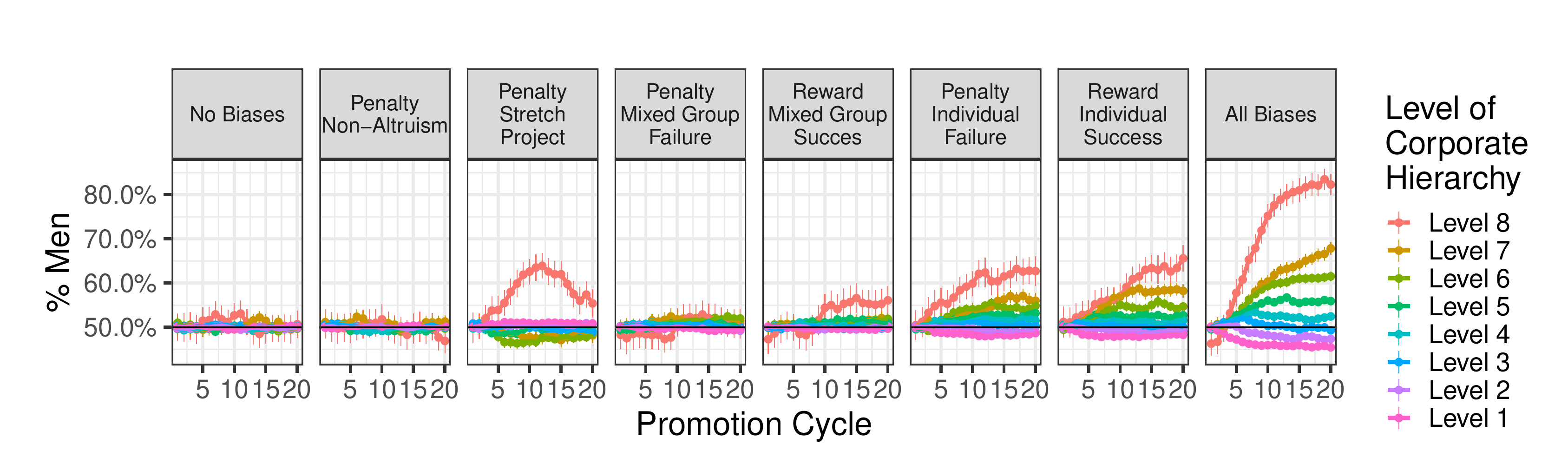}
         \caption{The y-axis represents the percentage of employees that are men}
         \label{fig:1a}
     \end{subfigure}
     \begin{subfigure}[b]{\textwidth}
         \centering
         \includegraphics[width=\textwidth]{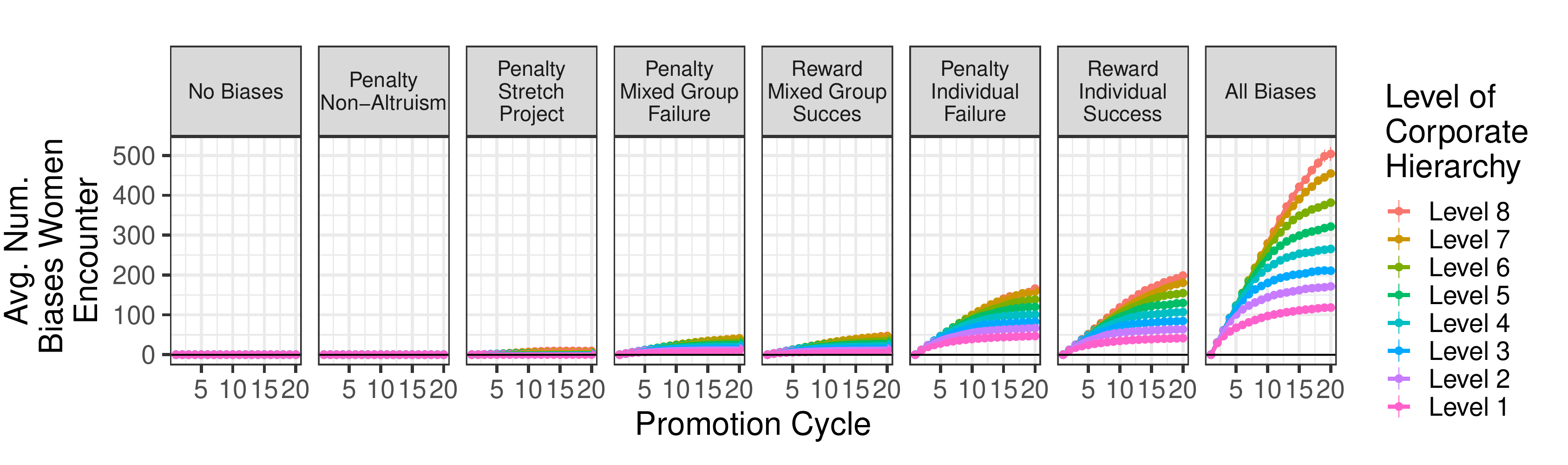}
         \caption{The y-axis represents the average number of biases that female agents encounter at corresponding levels of the corporate hierarchy}
         \label{fig:1b}
     \end{subfigure}

        \caption{Figures a) and b) differ only in their y-axis. both plots show results at each level of the corporate hierarchy (different colors) at each simulated promotion cycle (x-axis). Different subplots show results for simulations without any empirically-validated biases described in Table~\ref{tab:bias_list} (left-most), with all of these (right-most), or with each individually (middle sub-plots). Error bars represent confidence intervals from 100 randomly initialized simulation runs.}
        \label{fig:three graphs}
\end{figure}

Our first modification of the unbiased model introduces six gender biases that have significant empirical support: 
\begin{enumerate}
    \item \textbf{Women’s successes on projects are valued less than men’s} \citep{moss2012science,castillaGenderRaceMeritocracy2008,bowenEvaluatingGenderBiases2000,swimHeSkilledShe1996,swimJoanMcKayJohn1989,eaglyGenderEvaluationLeaders1992}. For instance, in a randomized double-blind study, \cite{moss2012science} found that when evaluating candidates for a lab manager position, science faculty at research institutions assigned lower competence values to female applicants than identical male applicants. 
    \item \textbf{Women’s errors and failures on projects are penalized more than men’s}. For instance, women surgeons experience greater decrease in referrals after a bad outcome: a male surgeon has to have three patient deaths to be penalized the way a female surgeon is penalized after one patient death \citep{sarsonsInterpretingSignalsLabor2017}. 
    \item \textbf{Women receive less credit in mixed-gender teams} \citep{sarsonsGenderDifferencesRecognition2021,sarsonsInterpretingSignalsLabor2017,heilmanNoCreditWhere2005}. For example, co-authoring a paper benefits women economists less than it does men: each co-authored paper increases men’s probability of achieving tenure 8.2\% but increases women’s probability of achieving tenure by 5.6\% \citep{sarsonsGenderDifferencesRecognition2021}.
    \item \textbf{Women receive more blame when a mixed-gender team fails}. \citep{egan2017harry,haynesWhoBlameAttributions2012} For example, when raters receive information about a group’s failure, they assign more blame to women \citep{haynesWhoBlameAttributions2012}.
    \item \textbf{Women are penalized for exhibiting non-altruistic behavior}. Likeability is important for promotions \citep{fanningInternalAuditorsUse2014} but research suggests that women are seen more unfavorably when they depart from behaviors considered to be stereotypically feminine. For example, women are seen as less likable when self-promoting \citep{rudmanSelfpromotionRiskFactor1998}, when using directive leadership styles \citep{rudmanPrescriptiveGenderStereotypes2001}, or simply when being successful in a male-stereotyped role \citep{heilmanWhyAreWomen2007a}.
    \item \textbf{Women receive fewer opportunities for growth}. Employees’ success is related to the opportunities they are given, particularly their access to “stretch” assignments that provide opportunities for learning and growth, access to decision makers, and experience in new career tracks \citep{wichertEarlyStretchAssignment2011,fernandez-araozCuriousCompetent2018}. Women often receive fewer assignments that allow them to develop new skills and report having less access to challenging assignments \citep{kingBenevolentSexismWork2012}. For example, the American Bar Association found that 44\% of women of color and 39\% of white women reported being passed over for desirable assignments in law firms, compared to two percent of white men \citep{rhode2017women}.
\end{enumerate}

These gender biases are empirically observed manifestations of more general, well-theorized social processes, including how women are stereotyped along affective dimensions of meaning \citep{fiskeModelOftenMixed2002,rogers_affective_2013}, and how women regulate their behaviors to align with cultural expectations \citep{hochschild1979emotion}. They have empirical support primarily at the interpersonal level; that is, they are enacted by way of one or more individuals evaluating another individual and then taking (or not taking) a particular action toward them. Our model assumes the same; that is, we assume that promotion decisions are made by individuals, and that those decisions are a function of bias accumulated via these six mechanisms.

Programmatically, these mechanisms are implemented into the unbiased model using the approaches described in Table~\ref{tab:bias_list}, and expanded upon in Section~\ref{sec:bias}.  Notably, the six mechanisms vary in their effects. In particular, biases in how stretch projects are allocated have significant impacts on the career of an individual, because stretch projects counts for three times as much as a typical project. This means that success on growth projects can rapidly drive individuals up the corporate hierarchy. In contrast, discounted rewards for women on projects have very small impacts.  Thus, a single instance of this form of bias, at a single point in time, directed at a single individual, has a minimal effect.

More specifically, to allocate credit for project success and failures, we first assume that the credit $c$ that an agent receives for a project is randomly drawn from a normal distribution. Two quantities are then of interest. First, how much more credit, on average, do men receive than women for success, and how much less do they lose than women upon failure? This quantity, while intuitive, is not easy to connect to existent studies of gender discrimination because it does not account for variance in the distribution of $c$. Consequently, following \cite{martellMalefemaleDifferencesComputer1996a}, our model parameters are also specified according to a second measure, the \emph{percentage of variance in credit received that is explained by gender}. We can then use results from prior empirical work; in particular, we rely like \cite{martellBiasExclusionMultilevel2012} on a meta-analysis from \cite{barrett1993american} that states gender accounts for approximately 1-5\% of the variance in hiring decisions. In our model, we fix a parameter $r^2$, which represents this variance quantity, to .022. This means we assume that gender explains approximately 2\% of the variation in credit allocation, about half of what \cite{barrett1993american} found in their study. This quantity translates to an "average bias" of 3\%; on average, men's perceived promotability increases (decreases) only 3\% more credit upon success (failure).

\begin{figure*}[t]

     \includegraphics[width=\textwidth]{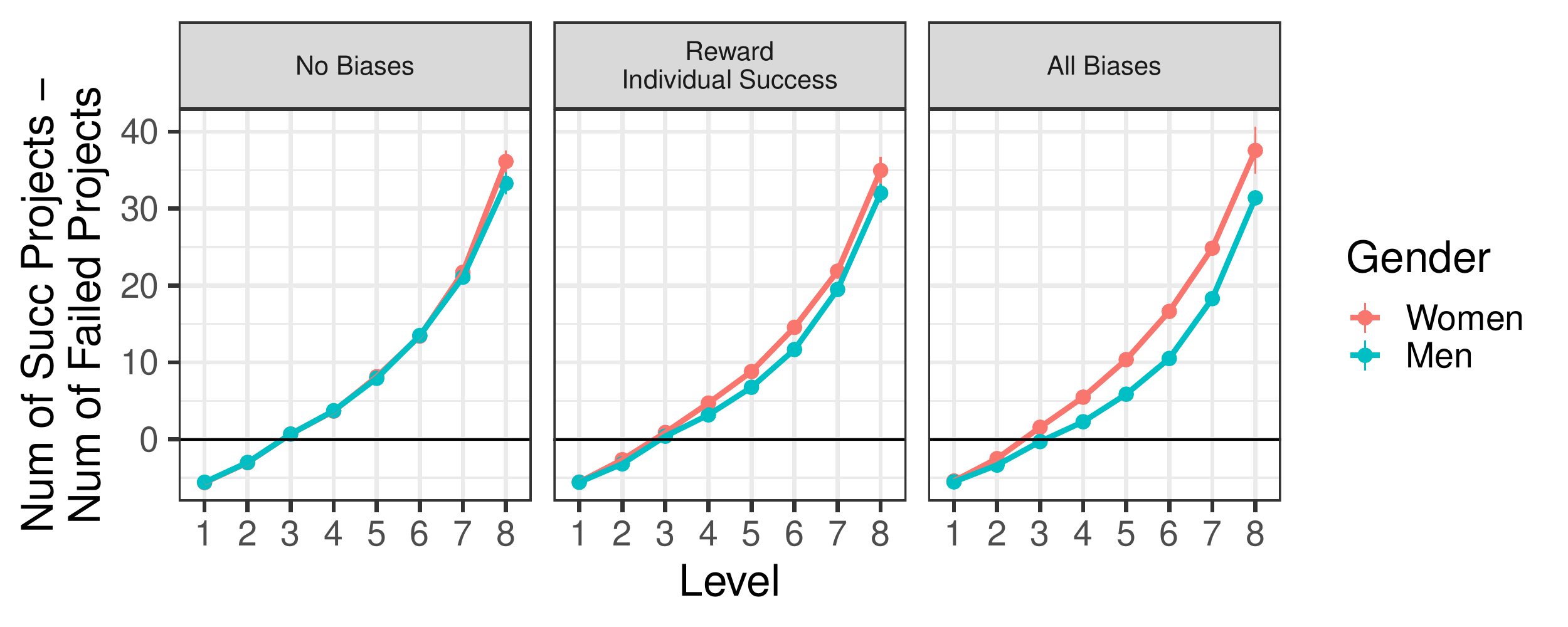}
    \caption{Difference between the number of successful projects and the number of failed projects (y-axis) for female (red) and male (cyan) agents at a given level of the corporate hierarchy (x-axis) for three different simulated conditions (separate subplots). These conditions are a subset of those used in Figure~\ref{fig:three graphs}. Values are calculated after 20 promotion cycles, and estimates and confidence intervals are constructed from 100 random simulation runs.}
    \label{fig:success}
\end{figure*}

Figure~\ref{fig:three graphs}a) shows that the interpersonal acts of discrimination we model lead to a glass ceiling effect. In the unbiased model, each level of the corporate hierarchy shows gender parity, with men and women both making up 50\% of the employees (left-most plot in Figure~\ref{fig:1a}). In contrast, with all of the mechanisms in Table~\ref{tab:bias_list} introduced into the model (right-most plot in Figure~\ref{fig:1a}), men dominate upper levels of the corporate hierarchy, leaving a preponderance of women at the lowest levels. 

Perhaps more interestingly, not all mechanisms we implement have the same impact. Rather, the most significant impacts come from mechanisms that are small but frequently applied. Figure~\ref{fig:1b} shows that the interpersonal acts of discrimination with the strongest effects on gender disparities were those that had \emph{most frequently been applied}, rather than those with the largest effects on individual agents. As an example of the latter case, differences in growth opportunities via stretch projects---which significantly alter career trajectories, but only for a small number of individuals--- impacted gender disparities at the top of the corporate ladder, but were too infrequent to reshape disparities at all levels.  Figure\ref{fig:1b} also suggests that women who reach high levels are affected more by devaluation for their successes than by penalties for failed projects. This result is explained by the fact that women at higher levels of the hierarchy are more successful (by chance, in our simulation) than women at lower levels.

Figure~\ref{fig:success} shows that women at high levels of the corporate hierarchy have a greater track record of successes than their male counterparts. In the unbiased model (left-most subplot), the difference between the number of successes and failures that employees had at different levels of the corporate hierarchy were the same for men and women. However, when interpersonal discrimination is introduced, women must have more success to achieve the same level of perceived promotability as men. Consequentially, women at the top of the hierarchy are, on average, significantly more successful than their male counterparts.

\subsection*{Incorporating Social Norms}

We have shown that enactment of gender bias at the interpersonal level, while having a potentially limited impact on a single individual at a single moment in time, can result, in aggregate, to a  glass ceiling for women. However, our model to this point does not express clear assumptions about \emph{why} interpersonal discrimination exists in the first place. Here, we provide such a mechanism based on the existence of gendered social norms at the meso and macro levels.

Our starting point is the empirical observation that fewer women in an environment correlates with increased gender discrimination. In management, in the Israeli army, among law students, and in blue-collar work groups, a greater proportion of men results in more bias against women \citep{lortie-lussierProportionWomenManagers2002,pazySexProportionPerformance2001,sackettTokenismPerformanceEvaluation1991,spanglerTokenWomenEmpirical1978}.  Prior work has expressed this empirical observation using a mathematical equation which purports that the degree of interpersonal discrimination at one level of the corporate hierarchy changes with the proportion of women at the level above \citep{robison-coxSimulatingGenderStratification2007}.  As gender disparities increase, then, gender discrimination does as well, rippling downward throughout the organization. 

However, this modeling assumption does little to address claims of "reverse discrimination". That is, such a model must either assume that gender imbalances that \emph{favor women} should result in discrimination \emph{against men}, or make the assumption that such reverse discrimination simply cannot exist. The latter claim is unsatisfying theoretically, because no underlying mechanism is suggested. But it is more consistent with reality. In the few settings where women dominate higher levels of the corporate hierarchy, there is little evidence of men’s promotion abilities being impacted. Instead, while women’s lack of representation in certain occupations exacerbates disadvantage,  men, namely heterosexual white men, when in short supply, enjoy a glass escalator, where they are put on a fast track to advanced positions \citep{budigMaleAdvantageGender2002,wingfieldRacializingGlassEscalator2009a}, and their evaluation is not affected by their proportion \citep{pazySexProportionPerformance2001}. The preponderance of male school superintendents is one such example.

Our model provides a mechanism that explains both how organizational gender disparities increase gender discrimination, and how this can apply only for women. To do so, we draw from scholarship on race and organizations, which emphasizes the importance of modeling both the meso- and macro-levels and how they are implicated in social inequality  \citep{rayTheoryRacializedOrganizations2019}. As such, we model the degree of interpersonal discrimination within an organization as a function of social norms that are both internal to the corporation (\emph{meso-level norms}) and external to the corporation at a societal level (\emph{macro-level norms}). We focus here only on project evaluations, but note that the model can easily be extended to other interpersonal biases we study as well. 

More specifically, we here introduce a mathematical model that defines the proportion of variance in project evaluations that is explained by gender as a function of social norms within and external to the simulated organization. Mathematically, our assumptions can be stated with the following pair of equations:

\begin{align}
r^2_{i} &= w \cdot B_{meso,i} + (1-w) \cdot B_{macro} \\
B_{meso,i} &= \frac{P_{i+1} - 0.5}{P_{m} - 0.5}  \cdot  B_{macro} 
\end{align}

Here, $r^2_{i}$ represents the proportion of variance that gender bias explains in project credit allocation \emph{at level $i$ of the corporate hierarchy}.  The parameter $r^2_{i}$ is a weighted sum of two quantities, where the weight $w$ is also a parameter of the model. The first quantity in the weighted sum is a macro-level norm $B_{macro}$. This parameter represents an assumption about the variance in project evaluations that would be explained by gender if social norms about gender were aligned only with societal expectations. The second is a meso-level norm $B_{meso,i}$, which represents the proportion of variance in project evaluations that would be explained by gender if norms were impacted by organizational structure. The value of $B_{meso,i}$ is determined via a formula consisting of $P_{i+1}$, the proportion of men at level $i+1$ at a given time in the simulation, and $P_{m}$, which represents a societal expectation of the percentage of men at a given level of the corporate hierarchy. This value is then multiplied by $B_{macro}$.

The value of $r^2_i$ encodes four core assumptions:
\begin{itemize}
    \item {\bf When the proportion of men in level $i+1$ of the company is 0.5,  $B_{meso,i}$ will be 0}: We assume that gender bias driven by social norms within the organization drops to 0 when gender equity is reached.
    \item {\bf When the proportion of men at level $i+1$ of the company is the same as the expected proportion given societal norms, $B_{meso,i}$ will represent the same value as the external norms $B_{macro}$.} We assume that $B_{macro}$ is an accumulation of norms from myriad gendered hierarchies across society. As such, a company with a hierarchy that matches societal expectations, $P_m$, should mirror the average societal norm, $B_{macro}$.
    \item {\bf The value of $r^2_i$ is a weighted average of meso-level and macro-level norms.} The model parameter $w$ encodes the modeler's belief about the relative importance of company structure-informed social norms as compared to societal expectations.
    \item {\bf Reverse discrimination can occur under certain conditions.}  $r^2_{i}$ can be negative, in which case we model reverse discrimination as in prior work. However, the parameters $w$ and $P_{m}$ can mitigate this possibility - if company structure matters little in comparison to societal expectations, or societal expectations are that a vast majority of men exist at a particular level of the corporate hierarchy, reverse discrimination is unlikely
\end{itemize}

\begin{figure*}[t]
    \centering
    \includegraphics[width=17.8cm]{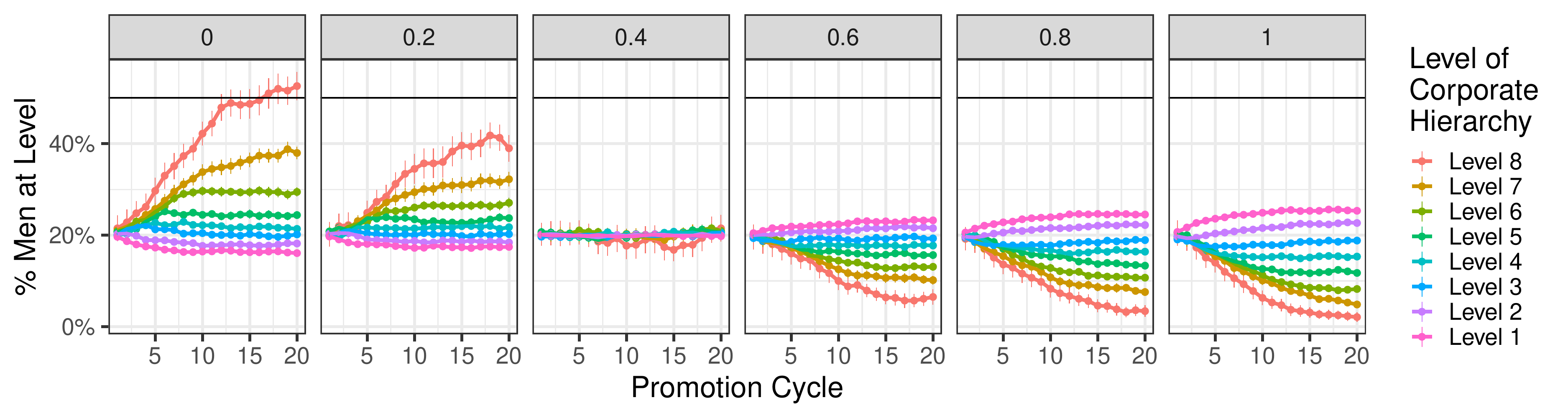}
    \caption{The percentage of employees that are men (y-axis) at each level of the corporate hierarchy (different colors) at each simulated promotion cycle (x-axis). Different subplots show results for simulations where we vary the strength of meso-level norms relative to macro-level norms (i.e. the model parameter $w$). A value of 0 represents a model in which only macro (societal) norms influence agent decisions, and 1 represents that only meso (organizational) norms impact agent decisions. All simulations here assume that, at the onset of the simulation, all levels of the corporate hierarchy are made up of 80\% women (i.e. that $P_{male}=.2$). Parameters used are otherwise the same as those in the All Biases condition displayed in Figure~\ref{fig:three graphs}. Error bars represent confidence intervals from 100 randomly initialized simulation runs. 
}
    \label{fig:norms}
\end{figure*}

We set $P_{m} = 0.7$, $B_{macro} = .01$, and then vary $w$ to explore the impact of weighting macro- vs. meso-level norms. As shown in Figure~\ref{fig:norms}, when we assume that macro-level norms have no influence (right-most subplot, $w=1$), men begin to face interpersonal discrimination in women-majority companies.  As these reverse biases are rarely observed empirically, we argue that a model that considers social norms only at the meso-level is incomplete. Instead, Figure~\ref{fig:norms} shows that only models which incorporate both meso- and macro- norms, and more specifically models that heavily weight societal-level norms relative to norms attributable to gender disparities within organizations, display evidence of the empirically observed glass escalator effect.

\subsection*{Implications for Intervention}

\begin{figure*}[h!]
    \centering
    \includegraphics[width=.85\textwidth]{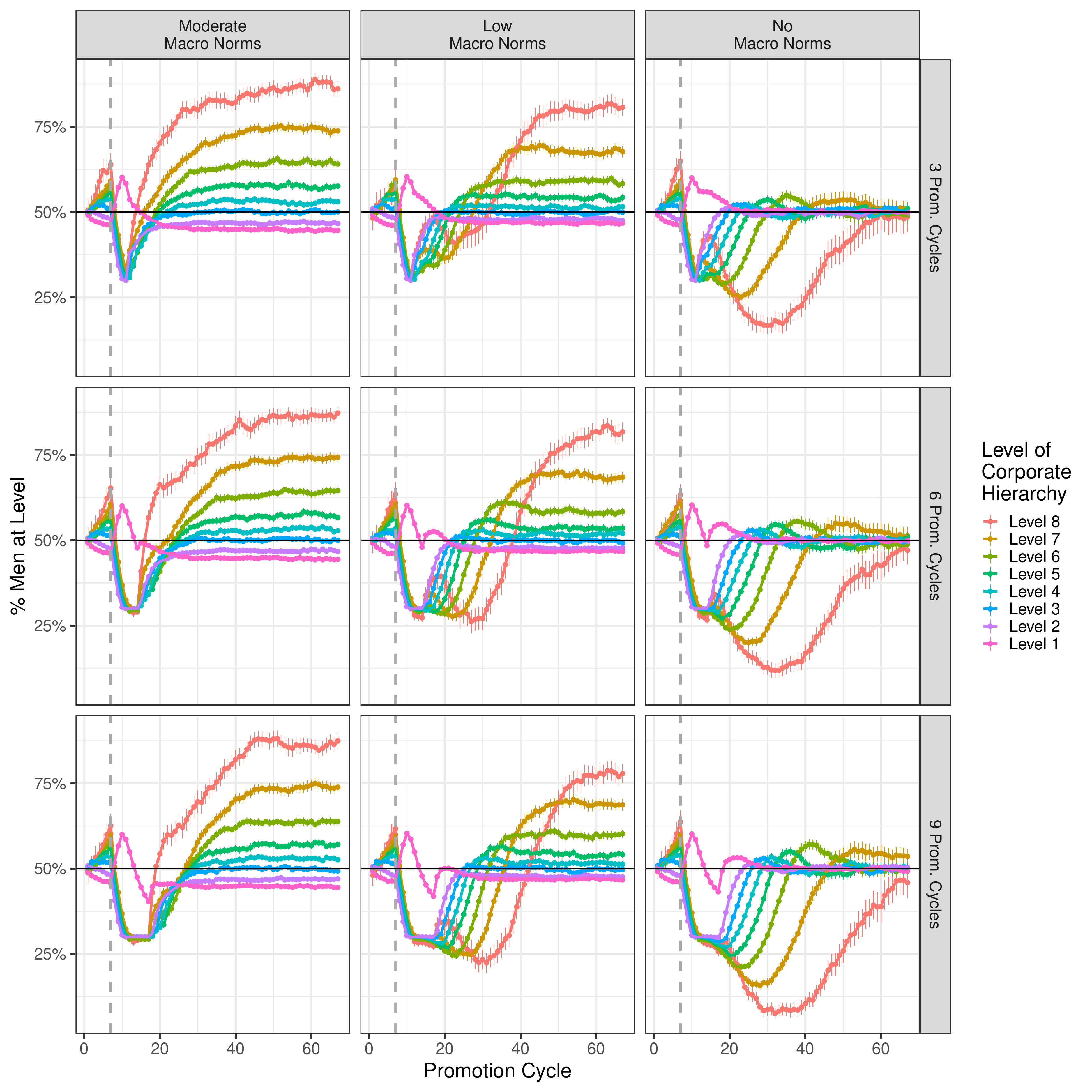}
    \caption{The percentage of employees at each level of the corporate hierarchy (different colors) that are men (y-axis) at each promotion cycle  (y-axis). Each column of sub-plots shows a different combination of weights for meso-level norms relative to macro-level norms. Each row considers a different number of promotion cycles over which the quota-based intervention is carried out. In all cases, the intervention begins after the first 7 promotion cycles. Error bars represent confidence intervals from 100 randomly initialized simulation runs.}
    \label{fig:intervention}
\end{figure*}

A common approach to mitigating gender disparities in organizations is to implement a quota-based system that enforces rules about promotions based on gender \citep{pandeGenderQuotasFemale2012}. Here, we simulate the effects of a quota-based intervention using our model. After 7 promotion cycles without intervention, and where interpersonal gender discrimination accounts for 1\% of the variation in project evaluations, a quota system is introduced to our simulated company. The quota intervention we assume is one where rules on promotions are enforced that target a goal of having 70\% of each level of the company above the entry level be women. We then vary 1) how long this intervention is carried out for, and 2) the assumed strength of meso-level norms, relative to macro norms, within the company. 

We find that regardless of the duration over which this intervention is conducted, gender disparities in our simulated organizations will return over time if company gender norms are at all displaced by gender-biased macro norms. These results are shown in Figure~\ref{fig:intervention}; only in the last row of figures, where macro-level norms have no impact on promotion decisions, do we observe lasting impacts of quota-based interventions.

These findings are similar to those by \cite{momennejadComputationalJusticeSimulating2019}, who show that changing structural conditions via varying the ratio of men to women within an organization without altering the culture created by previous gender discrimination will not address long-term inequality. However, our argument, and its implications, are distinct. Whereas \cite{momennejadComputationalJusticeSimulating2019} make the implicit claim that by changing company culture, one can reduce gender discrimination, our model argues that in addition to changing company culture, one must ensure that the company culture is strong enough to overpower macro-level gender norms. 

\section{Discussion}

To understand gender disparities within organizations, it is essential to understand them as the result of a complex, dynamic social system \citep{martellBiasExclusionMultilevel2012}. Prior agent-based models have shown how a variety of mechanisms, e.g. career interruptions and variable attrition rates \citep{robison-coxSimulatingGenderStratification2007,bullinariaAgentBasedModelsGender2018}, can create gender disparities in these \emph{complex organizational systems}. Most recently, \cite{momennejadComputationalJusticeSimulating2019} simulate the costs to individuals and institutions of sexist comments and objections to those comments in meetings, finding interrelationships between structural and learning effects. Their work shows how social learning can prevent structural interventions from being effective.

The present work extends these efforts in three important ways. First, we provide a concrete mechanism through which empirical observations of gender discrimination at the interpersonal level can be embedded into a model of complex organizational systems. Doing so paves a future path towards better integration of empirical and simulation results in the study of the glass ceiling. Second, prior work has largely focused on identifying the effect sizes of mechanisms for gender bias.  Our work instead models both effect size and the \emph{frequency} with which these small interpersonal acts of gender bias are enacted. This is important, because we find in our simulations that mechanisms of discrimination which produce small effects, but that occur at frequent intervals over a period of time, may be the most consequential in producing gender disparities.  

Finally, we introduce a new hierarchical model of how organizational and societal norms combined to create gender discrimination that is enacted within organizations. In doing so, we provide a new perspective on the impediments to interventions aimed at reducing gender disparities in organizations. Specifically, we argue via simulation that interventions aimed at reducing gender disparity in organizations must attend to the strength of societal gender norms and the stubbornness of outside influences when devising plans to disrupt gender homogeneity in corporate hierarchies. Critically, then, our model encourages further attention towards radical, societal-level change, or at least changes at the meso-level which can be expected to diffuse out to macro-level structures. An example of such efforts can be found in West Bengal, which reserved one third of village leadership positions for women. After two electoral cycles, quotas decreased men's stereotyping of women (as measured by the IAT), and the number of women who won positions that had not been reserved doubled \citep{beaman2009powerful}.

In considering these advancements of our work over prior efforts, however, it is important to also note our limitations. First, while we focus on \emph{perceived} gender, which is often---detrimentally---binary, we emphasize that gender itself is a continuous and socially constructed system \citep{ridgewayFramedGenderHow2011}.  Second, while our model could be easily extended to focus on intersectional dimensions of inequality and discrimination, the focus in the present work is on gender and thus does not account explicitly for the intersectional nature of inequality, or the ways in which stereotypes associated with other groups interact with gender stereotypes to amplify or dilute biases \citep{hallMOSAICModelStereotyping2019}. A survey of hundreds of women scientists, for instance, found that Asian American women reported the highest amount of backlash for self-promotion and assertive behavior \citep{williams2014double}.  Third, it is difficult to know the true impact of small continuously applied interpersonal biases. In any case, actual effect sizes will vary by organization, and by individuals within organizations. Our model, informed by empirical results, assumes very small effect sizes (on the order of 1 to 3\%); in the real world, these may be larger, smaller, or inconsistently applied. Fourth, there are many additional factors that contribute to any individual’s career trajectory beyond those listed here: choices and preferences, workplace family policies, and more. Our model demonstrates only that disparities on the order of magnitude of those seen in the real world can be achieved via the interpersonal mechanisms presented, with full account of the norms on which these interpersonal actions are based. More broadly, as with all agent-based models, we make few assumptions that the processes embedded within our model are a direct reflection of any one real company, but rather use the model to better understand the complexities of organizations \citep{epsteinWhyModel2008a}.

Despite these limitations, our work serves broader theoretical and policy-oriented goals. With respect to theory, our model provides a link between status construction theory \citep{markWhyNominalCharacteristics2009}, which focuses on the link between norms and behavior, with \citeauthor{rayTheoryRacializedOrganizations2019}'s [\citeyear{rayTheoryRacializedOrganizations2019}] theory of racial inequality emphasizing how culture, resources, and ideology interact  at the micro, meso, and macro levels of analysis. With respect to policy, laws are designed to address either large events that happen infrequently and can be easily attributed to a single actor—for example, overt sexual harassment by a manager-- or "pattern and practice" in an organization, for instance explicitly discriminatory policies. Our model shows, by contrast, how large organizational disparities can occur via that gradual and diffuse impact of many small, even unintentional events, decisions, and evaluations happening frequently over a long period of time. This raises important questions about the location of accountability within organizations and organizational culture, about what role the legal system or even workplace policies can or should play in cases where the biased evaluations are of the sort we model here.

\section{Materials and Methods}
We provide here more complete details on the simulation model used in this paper. Specifically, we first provide a more detailed overview of the model,  specifics on how the gender bias mechanisms are implemented, 
and how the quota intervention is implemented. Parameters settings are provided here inline. In the Appendix, we also provide a complete description of all model parameters, as well as tables that define the parameters used to generate specific figures in the main text. Full replication materials are available at \url{https://github.com/yuhaodu/workplace_gender_bias}.

\subsection{Model Details}\label{sec:models} In this section we provide more detail on the overall model implementation. 

\subsubsection{Agents}\label{sec:agents} As in all ABMs, agents in our model have state and can take actions. Agent states in our model are constituted by variables that keep track of the number of successful and failed projects this agent has completed, and the agent's perceived promotability. Each agent also has a binary attribute $a$ which represents their perceived binarized gender -- man or woman. 
When both male and female agents are initialized, they are seeded with an initial perceived promotability with a value drawn from a normal distribution $\mathcal{N}(\mu_{o},\sigma_{o})$. 

\subsubsection{Company}\label{sec:company} We model the same eight-level organization as \cite{martellMalefemaleDifferencesComputer1996a}. Level 8 represents the highest level of the company (i.e. the C-suite executives) and level 1 represents the lowest level. The number of positions at level $i$ are defined by the variable $N_i$. At the beginning of the simulation, all positions at all levels are seeded with agents, and the gender distribution is evenly split between the agents at all levels. In all simulations in this paper, as in  \cite{martellMalefemaleDifferencesComputer1996a}, the eight levels have 10, 40, 75, 100, 150, 200, 350, and 500 agents, respectively.

The company evolves through a series of $n_{sim}$ project turns. Each project turn can be either a traditional project turn or a stretch project turn. Stretch project turns occur once every $n_{stretch}$ turns, where here, $n_{stretch} = 12$, and $n_{sim}= 480$ unless otherwise noted. As such, the simulation iterates 20 times through a sequence of 11 traditional project turns, followed by one stretch project turn. On a traditional project turn, $P_{individual}$ percent of agents are randomly assigned to individual projects, and $1-P_{individual} $ percent of agents group projects. On a stretch project turn, stretch projects will first be assigned to $P_{stretch}$ percent of agents. Then $(1-P_{stretch}) * P_{individual}$ percentage of agents receive individual projects, while the rest will be assigned to group projects. In this work, $P_{stretch} = .1$ and $P_{individual} =.5$ for all runs.

After $n_{promotion}$ project turns, the company will carry out one \emph{promotion cycle turn}. Promotion cycle turns happen in a sequence of two steps. First, a random $P_{leave}$ percentage of agents at each level of the company leave (here, $P_{leave}=15\%$). Second, the company carries out a series of promotions, where empty positions caused by agents leaving the company are filled by agents who occupy the lower level positions.  Agents that are promoted are those that have the highest perceived promotability. Empty positions at lowest level are filled by new agents. 

\subsubsection{Project} \label{sec:proj} There are three kinds of projects in our simulation -- individual projects, stretch projects, and group projects. Individual projects and stretch projects are both assigned to a single agent. Group projects are assigned to two agents. All projects have an attribute, $c$, that is used to determine the amount of credit (blame) given to agents assigned to the project when it succeeds (fails). The value of $c$ is drawn from a normal distribution with mean $\mu_{r}$ and standard deviation $\sigma_{r}$ for individual and group projects, and from a normal distribution with mean $\mu_{st}$ and standard deviation $\sigma_{st}$ for stretch projects. Simulations in this paper are run with  $\mu_{r} = 10, \sigma_{r}=1, \mu_{st}=30, \sigma_{st} = 1$, reflecting an assumption of stretch projects being roughly three times as important as the typical project.

In our simulation, we make the simplifying assumption that all projects are equally likely to succeed or fail. With no gender bias, if a project succeeds, the perceived promotability of the agents assigned to the project will increase by $c$; if it fails, the perceived promotability of the agents assigned will decrease by $c$. 

\subsection{Mechanisms of Interpersonal Discrimination} \label{sec:bias}
Here, we provide additional details on the six mechanisms for gender bias introduced in the main text.

\subsubsection{Women’s successes on projects are valued less than men’s }\label{sec:reward_indiv}


We operationalize devalued success for women on projects using the percentage of variance in project credit that is explained by agent gender. More specifically, model parameters $r^2_i$, introduced in Equation [1], can be interpreted as the percent of variance explained by agent gender in a linear regression where the dependent variable is $c$, the credit the agent (at level $i$ of the company) receives for completing a successful project. In Figure~\ref{fig:three graphs}, credit received is independent of the agent's level of the company, and thus we discuss a parameter $r^2$, where $r^2_i = r^2 \forall i$. For Figure~\ref{fig:three graphs}, $r^2= .022$. Practically, this is implemented by setting $w$ to 0 in Equation [1] in the main text, and fixing $B_{macro} = .01$.

To to explain how gender bias in project credit allocation is implemented, we focus on this level-independent value $r^2$. The details stated here go through analogously with parameters $r^2_i$. Implementing percent variance explained in the simulation requires a variable transformation from percent of explained variance to a raw value, $d$, that differentiates credit given to women and credit given to men.  To translate from $r^2$ to $d$, we first expand notation, assuming the perceived promotability of a male agent will increase by  $c$ upon the completion of a successful individual project, while the perceived promotability of a female agent will increase by only $c-d$.  We then derive the appropriate value of $d$ such that this process will result in a particular value of $r^2$. 
To do so, note again that the credit of a project is drawn from a normal distribution with mean $\mu_r$ and $\sigma_r$. Now, define $ d = \frac{2 \cdot r}{\sqrt{1 - r^2}}$, such that $d$ represents the standardized mean difference between credit allocated to men and women.  Let us now define $\mu_{g}$ and $\sigma_{g}$ to represent the mean and standard deviation of project credit allocated to agents with gender $g$. Via simple derivation is can be said that $\mu_{male} - \mu_{female} = d \cdot \sqrt{2 \cdot (\sigma_{male}^2 + \sigma_{female}^2 )} $.  In turn, $\mu_{male} - \mu_{female} = d$ if we set the $\sigma_{r}$ to 1. 

Thus, by fixing $\sigma_{r} = 1$, as we do in the simulation, we can model the fact that gender explains $r^2$ percent of the variance in credit allocation via the following procedure. First, for a successful project, we sample credit $c$ for this project. The perceived promotability of a male employee will then increase by $c$, and the perceived promotability of a female employee will only increase by $c-d$. In this way, we can simulate an environment where gender bias accounts for $r^2$ proportion of the variance. 

The value $d$ is useful for another reason as well; the quantity $\frac{d}{c}$ can be understood as the average amount that a man's perceived promotability will increase over and above a woman's for the same successful project. That is, given fixed values for $r^2$, $\mu_r$, and $\sigma_r$, one can compare the raw percent increase that a male versus a female agent receives in perceived promotability for each successful project completed. Because of this dependence on some unknowable "absolute increase in promotability per project success", the quantity of interest for both our work and Martell et al. \citep{martellMalefemaleDifferencesComputer1996a} is thus not $\frac{d}{c}$ but $r^2$.

Finally, we note again that it is possible for $r^2_i$ to be negative. In this case, our simulation code instead models $\mu_{female} - \mu_{male} = d$, effectively encoding so-called "reverse discrimination".


\subsubsection{Women’s errors and failures on projects are penalized more than men's}\label{sec:penalty_indiv}

To model blame for failed projects, we adopt almost the same procedure as we do for credit for success. The only difference is that failed projects decrease perceived promotability, instead of increasing it. Analogously, men's perceived promotability decreases by $c$ for failed projects, and women's by $c+d$. 



\subsubsection{Women receive less credit in mixed-gender teams}\label{sec:reward_group}

To model biased allocation of credit in mixed-gender teams, we adopt the same procedure as we do for individual projects. The only difference is that we use a different parameter, $\widetilde{r}^2$, and analogously $\widetilde{r}^2_i$ when level-specific biases are considered. 


\subsubsection{Women receive more blame when a mixed-gender team fails}\label{sec:penalty_group} 
Again, for increased blame in mixed-gender teams, we adopt the same procedure as increased penalties for women in individual projects, but use the parameter $\widetilde{r}^2$ instead of $r^2$ to determine the strength of gender bias being applied. 


\subsubsection{Women are penalized for exhibiting non-altruistic behavior.}\label{sec:altruistic}

In our model, we assign a percent of women, $P_{com}$ to occasionally self-promote by complaining about unfairness when they experience bias. Doing so leads to a decrease their promotability score when they engage in this behavior by multiplying a discount factor $f_{dis}$ to their perceived promotability. If a female agent engages in self-promotion activity, their perceived promotability will change from $x$ to $f_{dis}\cdot x$ where $f_{dis} \leq 1$. In the simulations presented in the main text, $P_{com} = .1$, and $f_{dis}=.9$.


\subsubsection{Women receive fewer opportunities for growth.}\label{sec:growth}



In our model, at fixed intervals (every $n_{stretch} =5$ turns in the models in the present work),  we introduce stretch projects that provide outsized boosts in perceived promotability. Women need to achieve $P_{female}$ more successful projects than those of the average of qualified men to be assigned stretch projects. In the results presented here, $P_{female} = 20\%$; women thus need 20\% more successes to be considered for stretch projects. On each stretch project turn, we first rank the agents according their perceived promotability. The top $P_{stretch}$ percent of agents are then considered to be \emph{pre-qualified} for stretch projects. In the results presented here, $P_{strech} =10\%$. From these pre-qualified agents, we calculate the average number $n_{avg}$ of successful projects that male agents have already finished. Female agents then must have had to finish $n_{avg} \cdot (1 + P_{female})$ successful projects to be qualified for stretch projects.


\subsection{Modeling the Quota Intervention}\label{sec:quota}
The quota-based intervention study we introduce has a single parameter, $K$, that specifies a quota for the percentage of female agents expected at each level of the company.  Thus if level $i+1$ has $n$ positions, and $n_f$ is the number of female employees at level $i$, we will try to promote $n \cdot K\% - n_f$ female employees  from level $i$ to guarantee that there are at least $K\%$ female employees at level $i + 1$. Other positions at level $i+1$ are filled by employees who have highest perceived promotability from level $i$. 

We evaluate this intervention by further varying two additional parameters in Figure~\ref{fig:intervention}. Different rows of Figure~\ref{fig:intervention} refer to different ranges of project turns, $I_{range}$, which determine the project turn on which the intervention starts and the project turn on which the intervention ends. Values of $I_{range}$ [168,240],[168,312], and [168,384] correspond to the 4, 6, and 9 Promotion Cycles labels in the figure. We also vary the weight of meso-level norms. In all cases, the weight of meso-level norms starts with $w_0 =0$, i.e. norms are entirely determined by macro-level norms. Then, at the beginning of the intervention, and through the rest of the simulation, the weight will be altered to $w$. We set $w$ to 0.4, 0.7, and 1, aligning with the "Moderate Macro Norms", "Low Macro Norms", and "No Macro Norms" labels of the plot columns in Figure~\ref{fig:intervention}.

\section*{Acknowledgements}

We would like to thank Kathleen Carley for bringing a shared interest in gender discrimination and simulation to our attention four years ago. We also would like to thank Jonathan H. Morgan, Stefania Ionescu, and Cecilia Ridgeway for comments on prior versions of this article. We are also grateful to participants at the Networks 2021 conference and IC$^2$S$^2$ 2021, where this work was presented. Y.D. and K.J. were supported by Amazon and the NSF under
award NSF IIS-1939579.

\bibliographystyle{plainnat}
\bibliography{pnas-sample}

\appendix

\section{Description of Model Parameters}

\begin{table}[t]
\small
\begin{tabular} {|p{.1\textwidth}|p{0.7\textwidth}| p{0.2\textwidth}|} 
\toprule 
{\bf Parameter} & {\bf Explanation} & {\bf Default Value} \\ 
\toprule
\multicolumn{3}{|c|}{{\bf Model Constants}} \\ \midrule
$n_{sim}$ & The total number of project turns that are carried out  & 480\\
\hline
$P_{male}$ & The percentage of newly initialized agents that are men & .5\\ 
\hline
$N_{i}$ & The number of agents at level $i$ of the company & 500, 350, 200, 150, 100, 75, 40, 10 for L1-8\\
\hline
$n_{promotion}$ & The number of project turns before one promotion cycle turn occurs & 24  \\
\hline
$P_{leave}$ & The percentage of agents who leave the company on every promotion cycle turn & 15\% \\
\hline
$P_{s}$ & The probability any project is successful & 50\% \\
\hline
$\mu_{o},\sigma_{o}$&  The mean and standard deviation of the normal distribution from which initial promotability scores are drawn &  $\mu_{o} = 50 \:\:\: \sigma_{o} = 1$ \\
\hline
$\mu_{r},\sigma_{r}$ & The mean and standard deviation of the normal distribution from which credit is drawn to be distributed for an individual or group project & $\mu_{r} = 10 \:\:\: \sigma_{r} = 1$\\ 
\hline
$\mu_{st},\sigma_{st}$ &  The mean and standard deviation of the normal distribution from which credit is drawn for a stretch project & $\mu_{st} = 30 \:\:\: \sigma_{st} = 1$ \\ 
\hline
$P_{individual}$ & The percentage of the agents who receive individual (vs mixed) projects at project turn. & 50\% \\ 
\hline
$P_{stretch}$ & The percentage of the agents at each level of the simulated company that receive a stretch project & 10\% \\
\hline
$n_{stretch}$ & The number of project turns before one stretch project turn occurs & 12 \\ 
\hline
\midrule
\multicolumn{3}{|c|}{{\bf Relevant to Gender Bias Mechanisms}} \\ \midrule \\[-.7em]
$r^2 /\  \widetilde{r}^{2}$ & The percentage of variance that gender accounts for in individual/\ mixed-gender group project evaluation without the hierarchical norms model & 0.0 \\

\hline
$P_{female}$ &  The percent of the number of successful projects that female agent needs to finish more than that of male agent to be qualified for stretch projects & 0\% \\
\hline
$P_{com}$ & In a successful mixed gender project, the probability of woman who complains about unfair credit distribution compared to her male teammate. & 0\% \\
\hline
$f_{dis}$ & The discount factor that will be {multiplied to female agents' perceived promotability if they complain about unfairness they encounter} & 0\% \\
\hline
\midrule
\multicolumn{3}{|c|}{{\bf Relevant to Mechanism for Hierarchical Norms}} \\ \midrule
$B_{macro}$\,/ $\widetilde{B}_{macro}$  & The proportion of variance that gender bias explains in individual /\ mixed-gender group project credit allocation at macro level. & 0.01 \\ 
\hline
$P_{m}$  &  Societal expectation of the percentage of men at the given level of a typical organization (excluding Level 1) & 0.7 \\
\hline
$w$ & The weight of internal organizational norms about gender (meso-level norms), relative to societal (macro-level) norms & 0 \\\hline \\[-.7em]
$r^2_{i} /\  \widetilde{r}^{2}_{i}$ & The percentage of variance that gender bias accounts for in individual/\ mixed-gender group project evaluation at level $i$ of the company & - \\
\midrule 
\multicolumn{3}{|c|}{{\bf Relevant to Quota-Based Intervention}} \\ \midrule
$K$ & Quota of intervention on promotion which guarantees the percentage of female at each level except level 1. & 0 \\
\hline
$I_{range}$ & The range of simulation turns that intervention on promotion is carried out. & [0,0] \\ 
\hline
\end{tabular}
\caption{The left column shows the name of all parameters in our model. The center column provides a description of the meaning of the parameter. Finally, the right column shows the default value of the parameter for the unbiased model with no interventions. For clarity, parameters are separated by whether or not they are relevant to the model as a whole, to the six gender bias mechanisms we introduce, to our model for hierarchical, gendered social norms, or to the intervention.  Note that $r^2_{i} /\  \widetilde{r}^{2}_{i}$ does not have a default value, because it is determined from other parameters.}
\label{tab:parameter_name}
\end{table}

Table~\ref{tab:parameter_name} gives complete descriptions of all parameters used in the model described on our paper. The right-most column provides default values for a number of model parameters \emph{assuming the unbiased model} without any interventions.

\begin{table}[t]
\small
\begin{tabular} {|p{.3\textwidth}|p{0.7\textwidth}|} 
\hline
{\bf Named Conditions} & {\bf Parameters Changed from Default} \\ \hline \hline
No Biases & None, all defaults are used \\ \hline
Penalty Stretch Project &  $P_{female} = 20\%$ \\ \hline
Penalty Non-Altruism & $P_{com}=10\%$, $f_{dis}=90\%$   \\ \hline \\[-.7em]
Penalty Mixed Group Failure & $\widetilde{r^2}=.022$, only for failed projects \\ \hline \\[-.7em]
Reward Mixed Group Success &  $\widetilde{r^2}=.022$, only for successful projects \\ \hline \\[-.7em]
Penalty Individual Failure & $r^2=.022$, only for failed projects   \\ \hline \\[-.7em]
Reward Individual Success &  $r^2=.022$, only for successful projects  \\ \hline \\[-.7em]
All Biases &  $r^2=.022$ and $\widetilde{r^2}=.022$ for all projects, $P_{com}=10\%$, $f_{dis}=90\%$, $P_{female} = 20\%$  \\ \hline
\end{tabular}
\caption{Model parameters that are modified for the different named conditions in the main text}
\label{tab:params_for_first_part}
\end{table}

In the main text, we first experiment with different gender bias mechanisms. The named conditions used in Table~\ref{tab:bias_list}, and for Figure~\ref{fig:three graphs} and Figure~\ref{fig:success}, involve "turning on" different mechanisms of interpersonal gender discrimination.  Table~\ref{tab:params_for_first_part} shows which parameters change from the default values given in Table~\ref{tab:parameter_name} for each of the named conditions in the main text.

\begin{table}[t]
\small
\begin{tabular} {|p{.3\textwidth}|p{0.7\textwidth}|} 
\hline
{\bf Parameter} & {\bf Value} \\ \hline \hline
\\[-.8em]
$w$ & \{0,0.2,0.4,0.6,0.8,1\}, varied across the different subplots of Figure~\ref{fig:norms} \\
\hline
$B_{macro}$ & .01 \\
\hline
\\[-.8em]
$\widetilde{B}_{macro}$  & .01 \\
\hline
$P_{m}$ & 70\% \\
\hline
$P_{male}$ &  20\% \\ 
\hline
\end{tabular}
\caption{The specific parameters of the model of the experiment to test the effect of the weight of meso-level vs. macro-level norms in a female-dominated organization. Our experiment is carried out for each value in the curly bracket combined with values for other parameters.}
\label{tab:params_for_second_part}
\end{table}

We then introduce a model for hierarchical, gendered social norms. We provide experiments that describes how this model informs a theoretical explanation for the simultaneous existence of glass elevators and glass ceilings. Table~\ref{tab:params_for_second_part} shows which model parameters are changed from their default values for this experiment. Note that we focus in the paper only on the effect of hierarchical, gendered norms on project evaluations, although our model can be easily expanded to other mechanisms of interpersonal gender discrimination.

\begin{table}
\small
\begin{tabular} {|p{.3\textwidth}|p{0.7\textwidth}|} 
\hline
{\bf Parameter} & {\bf Value} \\ \hline \hline
$n_{sim}$ & 1600  \\
\hline
$B_{macro}$ & 1 \%  \\ 
\hline  \\[-.7em]
$\widetilde{B}_{macro}$ & 1 \% \\
\hline
$w_0$ & 0 \\
\hline
$w$ & \{0.4,0.7,1\} (Moderate, Low, No Macro Norms, respectively) \\
\hline
$K$ & 70  \\
\hline
$I_{range}$ & \{[168,240],[168,312],[168,384]\} (3, 6, 9 Prom. Cycles, respectively) \\
\hline
$P_{female}$ & 0 \% \\
\hline
\end{tabular}
\caption{The specific parameters of the model of the experiment to test the effect of a quota intervention on gender disparities. The experiment is carried out for each combination for values in the curly bracket combined with values for other parameters.}
\label{tab:params_for_third_part}
\end{table}

Finally, we conduct an experiment that analyzes the ramification for intervention of small but frequent acts of interpersonal gender discrimination that are driven in part by societal level norms. Parameters for these results are provided in Table~\ref{tab:params_for_third_part}.

\end{document}